\RequirePackage{fix-cm}
\documentclass[smallextended]{svjour3}       % onecolumn (second format)
\smartqed  % flush right qed marks, e.g. at end of proof
\usepackage{graphicx}
\usepackage{amssymb}
\begin{document}

\title{From change to spacetime: an Eleatic journey%\thanks{Grants or other notes
}

\titlerunning{From change to spacetime}        % if too long for running head

\author{Gustavo E. Romero}


\institute{Instituto Argentino de Radioastronom{\'{i}}a (IAR, CCT La Plata, CONICET) \at
              C.C. No. 5, 1894, Villa Elisa, Buenos Aires, Argentina. \\
              Tel.: +54-221-482-4903\\
              Fax: +54-221-425-4909\\
              \email{romero@iar-conicet.gov.ar}
}

\date{Received: date / Accepted: date}

\maketitle

\begin{abstract}

I present a formal ontological theory where the basic building blocks of the world can be either things or events. In any case, the result is a Parmenidean worldview where change is not a global property. What we understand by change manifests as asymmetries in the pattern of the world-lines that constitute 4-dimensional existents. I maintain that such a view is in accord with current scientific knowledge.

\end{abstract}

\begin{quotation}
\begin{flushright}
%$\varphi\'\upsilon\sigma\iota\zeta$ $\kappa\rho\'\upsilon\pi\tau\epsilon\sigma\theta\alpha\iota$ $\varphi\iota\lambda\epsilon\~\iota$.\\
Nature loves to hide. \\
{\sl Heraclitus, Fr. 123.}
\end{flushright}
\end{quotation}  

\section{Introduction}
\label{intro}

There is an essential tension threading the metaphysical discussion in the Western civilization along the last two and a half millennia. A tension between being and becoming, between substance and process, between things and events. Plato famously attributed to Heraclitus the doctrine that change is basic and that ``all things are in flux'' (DK 22A6)\footnote{See also Aristotle:  ``[Plato] as a young man became familiar with Cratylus and the Heraclitean doctrines that all sensible things are always flowing (undergoing Heraclitean flux)'' DK 65A3 (The notation refers to the doxography in H. Diels and W. Kranz, {\em Die Fragmente der Vorsokratiker}, 6th ed., Berlin, 1951.)}. However, as I have argued elsewhere (Romero 2012), there is nothing in the extant fragments of Heraclitus that may compel us to think that he denied substance and material things. Rather, on the contrary, the concept of material `thing' seems to make sense to Heraclitus only through change:

\begin{quotation}
By changing it is at rest. 
\begin{flushright}
(DK 22B84a)
\end{flushright}
\end{quotation}
   
Heraclitus, moreover, seems to share some ontological concerns with Parmenides, as shown in following fragment:

\begin{quotation}
[It] is wise to agree that all things are one. 
\begin{flushright}
(DK 22B123)
\end{flushright}
\end{quotation}  

It can be argued, on the basis of the extant fragments, that change is for Heraclitus a way to achieve the stability necessary for being (McKirahan 1994). If the waters of the river do not change, the river is not a river but a different thing, a lake. Some things can only be by changing. There is no necessary opposition between being and becoming, but rather the suggestion that becoming is the vehicle for being.  

In this paper I want to transit a similar path, going from change to being. In what follows I offer an ontological view where existent individuals can be things {\sl or} events, but in either case the ultimate reality is absolute being. I shall start building change upon things, following, in broad lines, a Bungean ontology (Bunge 1977, 1981). Then, I shall show how things can be construed as bundles of events. In both cases, I argue, the World is the totality of events, and such a totality is absolute and changeless being. I advocate for what William James, not without scorn, called the ``block universe''. I shall suggest that such a view is in accord with modern science. I shall certainly not be the first to maintain such position. The novelty is in the formal approach, that admits different ontic primitives and the assimilation, rather than the elimination, of becoming in a Parmenidean worldview.     

\section{Things}
\label{sec:1}

Individuals, whatever they are, associate with other individuals to yield new individuals. It follows that they satisfy a calculus, and that they are rigorously characterized only through the laws of such a calculus. These laws are set with the aim of reproducing the way real existents associate. Specifically, it can be postulated that every individual is an element of a set $s$ in such a way that the structure ${\cal S}=\left\langle s, \circ, \square \right\rangle$ is a \textit{commutative monoid of idempotents} 
(see Bunge 1977). This is a simple additive semi-group with neutral element.

In the structure ${\cal S}$, $s$ is the set of all individuals, the element $\square \in s$ is a fiction called the null individual (e.g. Martin 1965, Bunge 1966), and the binary operation $\circ$ is the association of individuals. Although ${\cal S}$ is a mathematical entity, the elements of $s$ are not, with the only exception of $\square$, which is a virtual individual introduced to form a calculus. The association of any element of $s$ with $\square$ yields the same element. The following definitions characterize the composition of individuals\footnote{This calculus of individuals differs from that of Leonard and Goodman (1940) in several aspects. Most notoriously, the inclusion of a virtual individual gives the set of all individuals a definite mathematical (above the logical) structure.}.

\begin{enumerate}

\item ${x} \in s $ is composed $ \Leftrightarrow  \left(\exists {y}, {z}\right)_{s_0} \left( {x} ={y} \circ {z} \right) $

\item ${x} \in s $ is simple $ \Leftrightarrow \; \neg \left(\exists {y}, {z}\right)_{s_0} \left({x} ={y} \circ {z} \right)$

\item $ {x}\subset {y}\ \Leftrightarrow {x} \circ {y} = {y}\ $ (${x}$ is part of ${y}\ \Leftrightarrow {x} \circ {y} = {y} $) 

\item $ \textsl{Comp}({x}) \equiv\{{y}\in s_0 \;|\; {y}\subset {x}\}$ is the composition of ${x}$.\\ 

\end{enumerate}

In definitions 1, 2 and 4, $s_0$ is $s-\{\square\}$. Individuals, so far, are ontologically neutral. We have introduced no specification of their nature.  

An individual with its properties make up a thing $X$. We can represent things through ordered pairs:
\[
	X=\left\langle x,{\cal P}(x)\right\rangle.
\]	
	
Here ${\cal P}(x)$ is the collection of properties of the individual $x$. A material thing is an individual with material properties, {\em i.e.} properties that can change (see below) in some respect.

Things are distinguished from abstract individuals because they have a number of properties in addition to their capability of association. These properties can be \textit{intrinsic} \normalfont (${\cal P}_{\rm i}$) or \textit{relational} \normalfont (${\cal P}_{\rm r}$). The intrinsic properties are inherent and they are represented by predicates or unary applications, whereas relational properties depend upon more than a single thing and are represented by $n$-ary predicates, with $n \geq 1$. Examples of intrinsic properties are electric charge and rest mass, whereas velocity of macroscopic bodies and volume are relational properties. Velocity (actually its modulus) is an intrinsic property only in the case of photons and other bosons that move at the speed of light in any reference system. 

Basic things can be easily introduced in this ontology as those things that are not composed by other things:

$$
X{\rm \; is\; basic\; iff\;} \neg(\exists Y)\; (Y \subset X), 
$$
or, equivalently, 
$$
\textsl{Comp}({X}) \equiv \emptyset.
$$

A thing $Z$ is said to be composed or complex, if:

$$
Z=X \circ Y \equiv \left\langle x \circ y, {\cal P}(x \circ y) \right\rangle,
$$
where $P(x \circ y)$ not necessarily satisfies ${\cal P}(x \circ y)= {\cal P}(x) \cup {\cal P}(y)$. If a property belongs to $Z$ but not to $X$ and $Y$, it is called an {\it emergent} property.

A fundamental methodological assumption of science is that properties can be represented by mathematical functions. For instance, temperature can be represented by a scalar field, velocity by a vector field, elasticity by a second rank tensor field, and so on. When we represent a given thing, we specify a set of functions that represent a collection of its properties.    

The {\em state} of a thing $X$ is a set of functions $S(X)$ from a 
domain of reference $M$ (a set that can be enumerable or nondenumerable) to the set of properties ${\cal P}(x)$. Every function in $S(X)$ represents a property in ${\cal P}(x)$. We postulate:

$$
(\forall P_{i})_{{\cal P}(x)} (\exists F_{i})_{S(X)} (F_{i}\hat{=}P_{i}).
$$

Here, the subindex $i$ runs over the different properties, $\hat{=}$ is the formal relation of representation (Bunge 1974), and the domain of the bound variables is made explicit.

Properties cannot change arbitrarily. There seems to be {\it restrictions} in the way they change. We can introduce such restrictions as {\it law statements}. These statements represent actual patterns of change in the world. Since properties are represented by functions, law statements are expressed by differential equations, or by integro-differential equations. Only functions that satisfy these equations can represent properties of existent (non-conceptual or material) things. Solutions for the equations can only be found if adequate boundary conditions are provided; hence the representation of any real thing must include its interaction with the environment, {\it i.e.} the rest of things.     

The set of the {\sl physically accessible} states of a thing $X$ is the {\em lawful state space} of
$X$: $S_{\rm L}(X)\subset S(X)$. The state of a thing is then represented by a point in $S_{\rm L}(X)$.

\section{Changes and events}
\label{sec:2}

 A change of a thing is any variation of its properties with respect to those of another thing\footnote{We call this second thing a {\it reference system.}}. In other words, change is the transition of a thing from one state to another. Only material things can change. Abstract things cannot change since they have only one state in any reference system (their properties are fixed by definition). 

An {\sl event} is a change of a thing $X$ with respect to a thing $Y$. Events can be represented by ordered pairs of states:

\[ e^{Y}_{X}=(s_1, s_2 ) \in E^{Y}_{\rm L}(X) = S_{\rm L}(X) \times S_{\rm L}(X). \]

The space $E^{Y}_{\rm L}(X)$ is called the {\em event space} of $X$ with respect to $Y$. A series of lawful events in $X$ , {\it i.e.} a continuous function over $E_{\rm L}(X)$ with respect to some reference thing, is a {\it process} in $X$. The totality of processes in a thing forms the {\it ontological history} of the thing. Let us call $h(X)$ the history of $X$ in some reference system. Then, 
$$  
(\forall X) (h(X)\subset E_{\rm L}(X)).
$$

If there are basic things, there could be basic events: $e^{Y}_{X}$ is basic iff $X$ is basic. For instance, the decay of a muon seems to be a basic event as far as we know. All processes, on the contrary, are composed events.

Following Bunge (1977, p. 225) we can introduce a partial order relation among events. Let $e$ and $e'$ be two events in a given event space of a thing $X$: $e\in E_{\rm L}(X)$ and $e'\in E_{\rm L}(X)$. Then, we say that $e$ {\it precedes} ($\prec$) $e'$ if $e\star e' \in E_{\rm L}(X)$. The operation of event composition $\star$ is defined by:

$$ 
e=(s_i, \; s_j) \in E_{\rm L}(X) \wedge e'=(s_l, \; s_m) \in E_{\rm L}(X) \rightarrow e \star e' = (s_i, \; s_m) \; {\rm only\; if} j=l.
$$
If $j\neq l$ the operation $\star$ is not defined. For any given thing $X$ and every associated event space $E_{\rm L}(X)$, $\left\langle E_{\rm L}(X), \prec \right\rangle$ is a strictly partially ordered set. Events in a thing $X$ are ordered as far as they have a common state. 

The \textit{Universe} $\cal{U}$ is the composition of all things:  $(\neg\exists X) \neg(X \subset \cal{U})$. Hence, the Universe is the maximal thing. As all things, the Universe has properties, and some of these properties can change respect to other properties of the same system\footnote{For instance, the energy density of the Universe can change with respect to its radius.}.  These changes determine the history of the Universe.

\section{Events as individuals}
\label{sec:3}

Both Russell (1914) and Whitehead (1929) formulated the program of considering events as basic individuals. Process philosophers often present things as ``processual complexes possessing a functional unity instead of substances individuated by a qualitative nature of some sort'' (Rescher 1996). Things, in this view, are construed as ``manifolds of processes''.  This project, however, has never been accomplished in a rigorous way and in accordance to modern science. There have been attempts to use the calculus of individuals of Leonard and Goodman (1940) to provide an outline of a formal ontology of events (e.g. Martin 1978), but the topological structure, based on the relation of precedence, attributed to the set of all events ($E$) is far too poor to account for some very general features of the world. Not all events can be related by a $\prec$-order. In particular, in vast systems, where events are not causally connected, the $\prec$-order can change with a change of reference system. More structure, in particular a metric structure, is required to deal with the totality of events. 

In what follows we shall consider events as individuals, and we shall develop an ontological view of things as derivative from events. This ontology should not be confused with the so-called `event calculus', originally proposed in logic programming form by Kowalski and Sergot (1986), which is a narrative-based formalism for reasoning about actions and changes.

We shall assume that the composition of events obeys that of general individuals:

\begin{enumerate}

\item An event ${e_{1}} \in E $ is composed $ \Leftrightarrow  \left(\exists {e_{2}}, {e_{3}}\right)_{E} \left( {e_{1}} ={e_{2}} \star {e_{3}} \right) $

\item An event ${e_{1}} \in s $ is basic $ \Leftrightarrow \; \neg \left(\exists {e_{2}}, {e_{3}}\right)_{E} \left({e_{1}} ={e_{2}} \star {e_{3}} \right)$

\item $ {e_{1}}\subset {e_{2}}\ \Leftrightarrow {e_{1}} \star {e_{2}} = {e_{2}}\ $ (${e_{1}}$ is part of ${e_{2}}\ \Leftrightarrow {e_{1}} \star {e_{2}} = {e_{2}} $) 

\item $ \textsl{Comp}({e}) \equiv\{{e_{i}}\in E \;|\; {e_{i}}\subset {e}\}$ is the composition of ${e}$.\\ 

\end{enumerate}

We can introduce a virtual null event $e^{0}$ stipulating that\footnote{Since the null event is a fiction, it has no ontological import. Its introduction allows to give to $\left\langle E, \star, e^{0}\right\rangle$ the same mathematical structure as adopted for a thing-based ontology.}: 

$$ 
(\forall e)_{E} (e^{0} \star e\equiv e).
$$

The composition of all events is the World ($W$):

$$
\neg(\exists e) \neg(e \subset W).
$$

The World, $W$, should not be confused with the Universe, ${\cal U}$, the composition of all things in a thing-based ontology as the one sketched in Sections \ref{sec:1} and \ref{sec:2}. The Universe can change, i.e. events and processes take place in the Universe. The World, the composition of all changes, can not change itself because it is not a thing. In an ontology of events, the totality of events is changeless, otherwise there would be a change not included in the totality, which is absurd. Events do not change, they {\it are} changes. In the sense used here for the words, the Universe can evolve, but not the World.      

We now need to introduce an ordering relation between events. The precedence relation $\prec$ defined before for the event space $E_{\rm} (X)$ of a thing $X$ is of no use now, since we have no things, and hence no states of things, to define such a relation. We cannot adopt neither a simple relation of ``before than'', as Carnap (1958), Gr\"unbaum (1973), and Martin (1978) did, because not all events can be ordered by such a relation without further specification: we know from relativity theory that such an order can be inverted by choosing an appropriate reference system in the case of space-like events. We need to introduce a stronger structure in the set of all events $E$, if we want to represent with such a set the World. To achieve this goal, we stipulate that $E$ is a {\it metric space}. \\

{\bf Definition}. A set $E$ is a metric space if for any two elements $e_{1}$ and $e_{2}$ of $E$, there is a real number $d(e_{1}, \; e_{2})$, called the {\it distance} between $e_{1}$ and $e_{2}$ in accordance with the postulates:\\

${\rm M1}. \;\; d(e_{1}, \; e_{2})=0 \;{\rm iff}\;e_{1} = e_{2}.$

${\rm M2}. \;\; d(e_{1}, \; e_{2})+d(e_{2}, \; e_{3})\geq d(e_{1}, \; e_{3}) \;{\rm with}\;e_{3} \in E.$\\

Lindenbaum (1928) has demonstrated that from these two axioms it follows that:\\

${\rm M3}. \;\; d(e_{1}, \; e_{2})=d(e_{2}, \; e_{1}).$
 
${\rm M4}. \;\; d(e_{1}, \; e_{2})\geq0.$\\

Now, only in case that  $d(e_{1}, \; e_{3})>0$, we can introduce a precedence relation between $e_{1}$ and $e_{3}$ :\\

{\bf Definition}.  The event $e_{1}$ {\it precedes} (or is {\it earlier than}) the event $e_{3}$ iff $(\exists e_{2})_{E} [d(e_{1}, \; e_{3})\geq d(e_{1}, \; e_{2})+d(e_{2}, \; e_{3})]$.\\

In short, $e_{1}\prec e_{3}$. Events such that $d>0$, $d=0$, and $d<0$ are called {\it time-like}, {\it null}, and {\it space-like} events, respectively. 

Given any event $e\in E$, we call the {\it future} of $e$ the set $F=\{ e' : d(e, e')>0 \wedge e \prec e' \}$. Similarly, we define the {\it past} of $e$ as the set $P=\{ e' : d(e, e')>0 \wedge \neg(e \prec e') \}$. Notice that every event has its own past and future, that depends on the metric of the space $E$. This contradicts the popular claim that the distinction between past and future requires consciousness. Rather on the contrary, there is no consciousness without memory, which in turn involves the
past-future distinction, in particular the distinction between the lived and the expected.

Once we have equipped the set of events with a metric structure, we can make the fundamental assumption of an event ontology: The World is represented by a metric space. In symbols:

$$  
E\hat{=}W.\\
$$
 
\noindent Here, $E$ is a mathematical construct and $W$ is the composition of all events, i.e. the maximal existent in an event ontology. 

A final step in the formulation of our event ontology is the formal construction of things out of events\footnote{Notice that I am providing here the foundations of an event ontology. I am not assuming a thing ontology in this section, so there is no semantic circularity. Rather: my goal is to show that event and thing ontologies are alternative representations of the same underlying reality. }. In order to achieve this we introduce the operation of abstraction from a collection of individuals. Let us consider a formula with a single variable $x$ that runs over events: `$(--x--)$'. This formula can be atomic or complex (build up out of atomic formulae connected by standard logic functors). The formula predicates of each individual $x$ such and such a property. We can abstract a virtual (i.e. fictitious) class from such a formula forming the collection (Martin 1969, p.125):

$$
P=\{ y: --y--\}.
$$       

Now, things can be constructed as classes of events sharing some properties, $P$, $Q$, etc:
$$
X=\left\langle P, \;Q, ...\right\rangle e.
$$

\noindent In this way things are bundles of events defined by shared properties, which are abstracted from conditions imposed on the events. The thing `Socrates', for instance, is a cluster of events sharing their occurrence in Greece, previous to such and such other events, including events like `talking with Plato', and so on. Once things are introduced in this way, we can deal with them as in the previous sections. If, instead, we consider things as basic, we can construct events out of states of things, and once we have a language rich enough as to have events as values of its variables, we can treat them as primitives. In this way, the ontic building blocks of our description of the World can be either things or events. Through both ways we arrive to the same picture of reality, as we shall see in the next section.

\section{The emergence of spacetime}
\label{sec:4}

The set of all events, $E$, is different from the set $E_{\rm L}(X)$ of physically possible events in a thing $X$ and from the set ${\cal E}$ of accessible events to all things. The latter is defined as:\\

$$ 
{\cal E}= \bigcup_{i} E_{\rm L}(X_{i}),\\
$$
where the index $i$ runs over all things. Most events in this set are virtual events that {\it never occur} in actual things; they are just lawful, possible pairs of states. The events that really occur are those that belong to the history of each thing. We can connect both ontological views, the thing-based and the event-based ontologies, through the following definitions:\\

$$  
E= \{e : e \in \bigcup_{i} h(X_{i})\}\\
$$

\noindent is the set of all actual changes and,

$$
W=h(\cal U),
$$

\noindent is the World. The World is the history of the Universe. In this way, independently of the primitive terms in our ontological basis (things or events) we arrive again to:

$$
E\hat{=}W.
$$

The mathematical model of the World can be improved imposing some additional constraints on the set $E$. To the metric postulates $M1$ and $M2$ we shall add the following postulates:\\

${\rm P1}.$ The set $E$ is a $C^{\infty}$ differentiable, 4-dimensional, real pseudo-Riemannian manifold.\\

${\rm P2}.$ The metric structure of $E$ is given by a tensor field of rank 2, $g_{ab}$, in such a way that the differential distance $ds$ between two events is: $ds^{2}=g_{ab} dx^{a} dx^{b}.$\\

A real 4-D manifold is a set that can be covered completely by subsets whose elements are in a one-to-one correspondence with subsets of $\Re^{4}$. The manifold is pseudo-Riemannian if the tangent space in each element is flat but not Euclidean. Each element of the manifold represents one (and only one) event. We adopt 4 dimensions because it seems enough to give 4 real numbers to provide the minimal characterization of an event. We can always provide a set of 4 real numbers for every event, and this can be done independently of the intrinsic geometry of the manifold. If there is more than a single characterization of an event, we can always find a transformation law between the different coordinate systems. This is a basic property of manifolds. 

We introduce now the Equivalence Principle and the characterization of the metric through two additional postulates:\\

${\rm P3}.$ The tangent space of $E$ at any point is Minkowskian, i.e. its metric is given by a symmetric tensor $\eta_{ab}$ of rank 2 and trace $-2$.\\

${\rm P4}.$ The metric of $E$ is determined by a rank 2 tensor field $T_{ab}$ though the Einstein field equations:

\begin{equation}
G_{ab}-g_{ab}\Lambda=\kappa T_{ab}.\\
\end{equation}

In these equations $G_{ab}$ is the so-called Einstein tensor, formed by second order derivatives of the metric. In the second term on the left, $\Lambda$ is called the cosmological constant, whose value is thought to be small but not null. The constant $\kappa$ on the right side is $- 8\pi$ in units of $c=G=1$. Finally, $T_{ab}$ represents the source of the metric field, and satisfies conservation conditions ($ \nabla_b T^{ab}  \,  = T^{ab}{}_{;b}  \, = 0$) from which the equations of motion of physical things can be derived. The solutions of such equations are the histories of things whose properties are characterized by $T_{ab}$. Alternatively, the solutions can be seen as continuous series of events represented on the manifold $E$. The Einstein field equations are a set of ten non-linear partial differential equations for the metric coefficients.

\section{Change as asymmetry in a Parmenidean world}
\label{sec:5}

Postulates P1 to P4 given in the previous section, with an adequate formal background (Bunge 1967, Covarrubias 1993, Perez Bergliaffa et al. 1998), imply the theory of general relativity. What we have called the `World', in relativity theory is known as `spacetime' (${\cal ST}$). The representation of spacetime is given by a 4-dimensional manifold equipped by a metric. In standard relativistic notation:

$$  
{\cal ST}\hat{=}\left\langle E, g_{ab}\right\rangle.
$$

We remark that spacetime {is not} a manifold (i.e. a mathematical construct) but the ``totality'' (the composition in our characterization) of all events. A specific model of the World requires the specification of the source of the metric field. As we have seen, this is done through another field, called the ``energy-momentum'' tensor field $T_{ab}$ (Hawking and Ellis 1973). Hence, a model of the World is:

$$  
M_{\rm W}=\left\langle E, g_{ab}, T_{ab}\right\rangle.
$$

Since the ontic basis of the model is the {\it totality} of events, the World is ontologically determined. This does not imply that the World is necessarily {\it predictable} from the model. In fact, Cauchy horizons can appear in the manifold $E$ for many prescriptions of $T_{ab}$ (e.g. Joshi 1993). One thing is the World, and another our representations of the World.

In the World, objects are 4-dimensional bundles of events (Heller 1990). Beginning and end, are just boundaries of objects, in the same way that the surfaces and boundary layers are limits of 3-dimensional slices of such objects. The child I was, long time ago, is just a temporal part of me. The fact that these parts are not identical is not mysterious or particularly puzzling, since spacetime, although changeless itself, is composed of changes. We can understand such changes as asymmetries in the geometry of spacetime. We can quantify the intrinsic change rate of spacetime using the Raychaudhuri's equation (Raychaudhuri 1955).

Let us consider a time-like vector field $u^{a}$ that is tangent to the geodesics (basic processes or histories of basic things) of a spacetime ${\cal ST}$. We can define an {\it expansion scalar} $\theta=\nabla_{a}u^{a}=u^{a}_{;a}$, such that if $\theta>0$ the geodesics fly apart, if $\theta<0$ the geodesics come together, and if $\theta=0$ they remain self-similar. The Raychaudhuri's equation provides the evolution of $\theta$ with the separation between events:

$$
\frac{d\theta}{ds}=-\frac{1}{3} \theta^{2} - \sigma + \omega - R_{ab}u^{a}u^{b}.
$$           

\noindent In this equation $R_{ab}$ is the so-called Ricci tensor, that is formed with the second order derivatives of the metric $g_{ab}$, and $\sigma$ and $\omega$ are parameters that measure the {\it shear} and {\it rotation} of the geodesics, respectively. The derivative $d\theta/ds$ gives a measure of the rate of change in the history of the Universe $h({\cal U})$. If $\theta=0$ all slices of the World are identical. Only if $\theta\neq 0$ there is change from one slice to another, and we can say that the Universe evolves (undergoes change), or, what is the same, that the World presents asymmetries. Change, then, is an internal relative feature of the World, which, from a global point of view, is changeless. In Heraclitus words: ``By changing it remains at rest''. 

\section{Conclusions}
\label{sec:6}

In this paper I have outlined a formal ontological system that can accommodate either things or events as primitive individuals. The result, in any case, is a Parmenidean view of the World, where change is not possible for the totality of the existents. Nonetheless, change plays a fundamental role in the constitution of the World, as Heraclitus pointed out. In this sense, this work expands and elaborates on a central theme of Western metaphysics. I remark that the ontological views presented here are independent of the nature (quantum or not) of the basic building blocks of the World. An ontology is the most general theory that can be presented about what there is. It should be wide enough to accommodate all results of the factual sciences and it should provide a framework to stimulate further research. As any theory, an ontology should be testable. Not directly against experiment, but through its cogency with the totality of our scientific knowledge. It is my hope that the considerations presented in this paper can be useful as a contribution to a science-oriented metaphysics, rooted in a tradition initiated by Heraclitus and the Eleatic challenge to the Ionian concept of change.   

\section*{Appendix: Metric in the quantum world}
\label{Appendix}

It might be thought that in a quantum mechanical description of the world, the line element cannot be defined because such a description does not contain `particle' coordinates, and hence there are not sharp points or lines. This, however, is based on a misunderstanding of the metric concept of space-time. Space-time is formed out of {\em events}, not of physical objects such as elementary particles. The events are ordered pairs of {\em states}, in the framework of a thing-based ontology. States in quantum mechanics are represented by rays in a rigged Hilbert space. Actually, it is a basic postulate of quantum mechanics that every quantum system and its environment have an associated Hilbert space (Bunge 1967; Perez-Bergliaffa et al. 1993, 1996). Changes or events in quantum mechanics are pairs of rays in the Hilbert space of the system. The fact that the properties of quantum things only have probabilities (or propensities) of taking sharp values does not preclude change, and hence space-time, of the quantum description of nature. 

Moreover, quantum mechanics and quantum field theory presuppose space-time (e.g. Perez-Bergliaffa et al. 1996, axioms $A_{1}-A_{5}$). The usual presentation of these theories is in a flat, Minkowskian space-time, although they can be applied with certain caution to curved space-times (see, for instance, Wald 1994). Metric problems can appear only when change cannot be properly defined, as seems to occur around the Planck scale. At this scale, the continuum hypothesis breaks, and quantum gravity should replace metric gravity (e.g. Rovelli 2004). But this is neither the time nor the place to discuss the ontological implications of quantum gravity for the theory of change. {\em I'll be back} on this issue elsewhere.

\begin{acknowledgements}
I thank Mario Bunge, Santiago E. Perez-Bergliaffa, V. Bosch-Ramon, and Daniela P\'erez for stimulating discussions.
\end{acknowledgements}


\bibliographystyle{aipproc}   % if natbib is available


\newpage

\section*{Gustavo E. Romero} Full Professor of Relativistic Astrophysics at the University of La Plata and Chief Researcher of the National Research Council of Argentina. A former President of the Argentine Astronomical Society, he has published more than 250 papers on astrophysics, gravitation, and the foundation of physics, and 8 books. His main current interest is on black hole physics and ontological problems of space-time theories.

\end{document}